\documentclass[12pt]{emulateapj}

\shorttitle{Distance to the MCs} 
\shortauthors{Inno et al.}

\begin{document}
\title{On the distance of the Magellanic Clouds using 
Cepheid NIR and optical--NIR Period--Wesenheit Relations}  

\author{
L.~Inno\altaffilmark{1,2},
N.~Matsunaga\altaffilmark{3},
G.~Bono\altaffilmark{1,4},
F.~Caputo\altaffilmark{4},
R.~Buonanno\altaffilmark{1,5},
K.~Genovali\altaffilmark{1},
C.D.~Laney\altaffilmark{6,7},
M.~Marconi\altaffilmark{8},
A.M.~Piersimoni\altaffilmark{5},
F.~Primas\altaffilmark{2}, 
M.~Romaniello\altaffilmark{2}
}

\altaffiltext{1}{Dipartimento di Fisica, Universit\`a di Roma Tor Vergata, 
via della Ricerca Scientifica 1, I-00133 Rome, Italy; laura.inno@roma2.infn.it}
\altaffiltext{2}{European Southern Observatory, Karl-Schwarzschild-Str. 2,D-85748 Garching bei Munchen, Germany}
\altaffiltext{3}{Kiso Observatory, Institute of Astronomy, School of Science, The University of Tokyo
 10762-30, Mitake, Kiso-machi, Kiso-gun,3 Nagano 97-0101, Japan}
\altaffiltext{4}{INAF--OAR, via Frascati 33, I-00040 Monte Porzio Catone, Rome, Italy}
\altaffiltext{5}{INAF-Osservatorio Astronomico di Collurania, via M. Maggini,I-64100 Teramo, Italy}
\altaffiltext{6}{Department of Physics and Astronomy, N283 ESC, Brigham Young University, Provo, UT 84601, USA}
\altaffiltext{7}{South African Astronomical Observatory, P.O. Box 9, Observatory 7935, South Africa}
\altaffiltext{8}{INAF-Osservatorio Astronomico di Capodimonte, via Moiarello 16,I-80131 Napoli, Italy}

\date{\centering drafted \today\ / Received / Accepted }

\begin{abstract}
We present the largest near-infrared (NIR) data sets, $JHK_{\rm S}$, ever
collected for classical Cepheids in the Magellanic Clouds (MCs). We
selected fundamental (FU) and first overtone (FO) pulsators, and
found 4150 (2571~FU, 1579~FO) Cepheids for Small Magellanic Cloud
(SMC) and 3042 (1840~FU, 1202~FO) for Large Magellanic Cloud (LMC).
Current sample is 2--3 times larger than any sample used in previous
investigations with NIR photometry. We also discuss optical $VI$
photometry from OGLE-III. NIR and optical--NIR Period-Wesenheit
(PW) relations are linear over the entire
period range ($0.0<\log P_{\rm FU} \le1.65 $) and their slopes are,
within the intrinsic dispersions, common between the MCs. These are
consistent with recent results from pulsation models and
observations suggesting that the PW relations are minimally affected
by the metal content. The new FU and FO PW relations were calibrated using a
sample of Galactic Cepheids with distances based on trigonometric parallaxes 
and Cepheid pulsation models. 
By using FU Cepheids we found a true distance moduli of 
$18.45\pm0.02{\rm(random)}\pm0.10{\rm(systematic)}$~mag (LMC) and
$18.93\pm0.02{\rm(random)}\pm0.10{\rm(systematic)}$~mag (SMC).
These estimates are the weighted mean over 10 PW relations and
the systematic errors account for uncertainties in the
zero--point and in the reddening law.
We found similar distances using FO Cepheids
($18.60\pm0.03{\rm(random)}\pm0.10{\rm(systematic)}$~mag [LMC] and
$19.12\pm0.03{\rm(random)}\pm0.10{\rm(systematic)}$~mag [SMC]).
These new MC distances lead to the relative distance,
$\Delta\mu=0.48\pm0.03$~mag (FU, $\log P=1$) and
$\Delta\mu=0.52\pm0.03$~mag (FO, $\log P=0.5$),which agrees quite well
with previous estimates based on robust distance indicators.
\end{abstract}

\keywords{ Magellanic Clouds ---  stars: variables: Cepheids --- 
stars: distances --- stars: oscillations}

\maketitle

\section{Introduction}

Recent detailed investigations indicate that 2\%--3\%   
of the systematic error affecting the Hubble constant estimate is due 
to the Cepheid distance to the Large Magellanic Cloud \citep[LMC][]{madore10,riess11, freedman12}.
Moreover, the Magellanic Clouds (MCs) are fundamental benchmarks to constrain 
the accuracy and the precision of the most popular primary distance indicators 
\citep[][2011]{pietrzynski10,matsunaga09}.
The decrease of a factor two in metallicity between the LMC and the 
Small Magellanic Cloud (SMC) makes these galaxies also excellent laboratories 
to constrain the possible dependence of different standard candles on the metal 
content. Although, the MC Cepheids play a crucial role in many astrophysical 
problems, the number of homogeneous optical (\textit{B,V,R,I}) and near-infrared 
(NIR; \textit{J,H,K$_{\rm S}$}) data sets is quite limited. 

The most extensive surveys in 
the optical bands (\textit{V,I}) were performed by micro--lensing experiments 
(MACHO, EROS, OGLE).
The MACHO project collected $R$, $I$ band data for $\sim$1900 Cepheids in the 
LMC \citep{skrutskie06,allsman00,welch96}, 
while EROS collected $V,R$ band data for $\sim$300 and $\sim$600 Cepheids in 
the LMC and in the SMC, respectively \citep{marquette00}.
The most complete sample of MC Cepheids was collected by OGLE-III \citep[][2010]{sos2008}.
Their catalog includes  $V,I$ band light curves for more $\sim$7000 Cepheids 
(LMC: 2000 fundamental [FU], 1000 first overtone [FO];  
SMC: 2500 FU, 1500 FO). 
Accurate distance determinations to the MCs based on optical Period--Wesenheit (PW) 
relations have also been provided by \citet{udalski99}, \citet{bono02}, \citet{greo03} and  \citet{ngeow09}.
More recently, \citet{dicriscienzo12} provided a detailed theoretical investigation concerning the PW relations 
in the Sloan Digital Sky Survey bands.

The NIR data bases for MC Cepheids are significantly smaller:  \citet[][1994]{laney86} 
collected NIR light curves for 44 MC Cepheids (21 LMC, 23 SMC), while 
\citet{welch97} for 91 SMC Cepheids. More recently \citet[][hereinafter P04]{persson04} 
collected NIR light curves for 92 LMC Cepheids.
More recently, accurate $K$ band photometry (12 phase points per
variable) was
collected by \citet[][hereinafter R12]{ripepi12} in two LMC fields located around 30 Doradus
(172 FU, 152 FO) and the South Elliptical Pole (11 Cepheids). They
provided, by using also literature data, accurate estimates of Period--Luminosity (PL), PW, and
Period-Luminosity-Color relations for both FU and FO Cepheids.
One of the key advantages in using NIR data is that the pulsation amplitude 
decreases for increasing wavelength and the estimate of the mean magnitude becomes easier.
The previous largest set of single epoch measurements for MC Cepheids was collected 
by \citet{groenewegen00} (LMC: 713 FU, 450 FO; SMC: 1200 FU, 675 FO) using 2MASS 
and DENIS data sets.  
The same approach was also followed by \citet{nikolaev04} using MACHO and 
2MASS data sets (LMC: 1357 FU, 749 FO) and more recently by \citet{ngeow09} 
using the 2MASS data set (LMC: 1761 FU) and the mid-infrared SAGE \textit{Spitzer}
\citep{meixner06} data set (LMC: 1759 FU).

In this investigation, we provide new MC distances using a new sample of single-epoch 
\textit{J,H,K$_{\rm S}$} measurements of a significant fraction of MC Cepheids ($\sim$80\%) 
detected by OGLE. In  particular, in Section~2 we discuss the NIR and optical data sets 
we adopted in this investigation together with the criteria to select both FU and FO Cepheids. 
In Section~3 we present the PW relations, while  in Section~3.1 we focus our attention on the linearity 
and the metallicity dependence of NIR and optical-NIR PW relations. Empirical and 
theoretical absolute calibrations of the PW relations are addressed in Section~3.2. 
The summary of the results and more detailed discussion concerning pros and cons 
of the two independent calibrations are given in Section~4, while in Section~5 we briefly 
outline some possible future avenues concerning the developments of this project.

%
\section{Data sets and data selection}

The Cepheid intrinsic parameters are taken from the OGLE-III 
catalog \citep[][2010]{sos2008}. We adopted the following Cepheid parameters: 
pulsation period, position (right ascension and declination), mean $V$ and $I$ band
magnitude, $I$ band amplitude, epoch of maximum and pulsation mode. 
The optical OGLE-III Cepheid catalog was cross-correlated with the NIR catalog 
of the IRSF/SIRIUS Near-Infrared Magellanic Clouds Survey provided by \citet{Kato07}.
The single-epoch \textit{J,H,K$_{\rm S}$} magnitudes for the OGLE-III FU Cepheids were 
extracted by \citet{matsunaga11}. In this investigation we also included FO Cepheids. 
We ended up with a sample of 3042 LMC (1840 FU, 1202 FO) and 
4150 SMC (2571 FU, 1579 FO) Cepheids with three NIR (\textit{J,H,K$_{\rm S}$}) 
single-epoch measurements. 
The IRSF/SIRIUS \textit{J,H,K$_{\rm S}$} measurements were transformed into the 2MASS NIR 
photometric system following \citet{Kato07}.

The mean magnitudes of FU Cepheids were estimated using the NIR template light 
curves provided by \citet{templ}. 
To assess the accuracy of this method, we compared our estimates
of mean magnitudes with the mean magnitudes 
for LMC Cepheids based on finely sampled light curves (P04). 
The template light curves and the mean magnitude by P04 were also transformed 
into the 2MASS NIR photometric system following \citet{carpenter01}. 
Figure~1 shows that the intrinsic dispersion decreases by a 
factor of two when moving from the single-epoch measurements to the mean magnitudes 
based on the template (0.12 versus 0.05 mag).
The total error budget of the mean magnitudes estimated using the template 
light curve is given by 
$\sigma^{2}_{\lambda_i}=\sigma^{2}_{m_i}+\sigma^{2}_{\rm cal}+\sigma^{2}_{\rm rph}$, 
where $\sigma^{2}_{m_i}$ is the intrinsic photometric error, with a typical 
value of $\sim 0.03$ at 16 mag in \textit{J,H,K$_{\rm S}$}; 
$\sigma^{2}_{\rm cal}$ is the error due to the transformation into the 2MASS 
photometric system, it is of the order of 0.01 mag for UKIRT, LCO and 
IRSF systems;  
$\sigma^{2}_{\rm rph}$ is the scatter due to the random phase sampling, 
it is given by the template algorithm and it is $\sim$0.05 mag \citep{templ}.

For the FO Cepheids the mean magnitude is based on the single epoch 
measurements, since template light curves are not available for these 
pulsators. It is worth mentioning, that the mean magnitudes of these pulsators 
are less affected by their random sampling, since their luminosity amplitude 
is on average three times smaller than for FU Cepheids \citep{madfreed10}.  
The errors of the FO mean magnitudes were estimated using the above relation, 
but the term $\sigma^{2}_{\rm rph}$ gives the typical semi-amplitude of FO light 
curves ($\sim$0.10 mag).
We plan to address the discussion concerning the template light curve for FO 
Cepheids and their errors in a forthcoming paper.

\section{Period Wesenheit relations}
The Wesenheit indices, introduced by \citet{madore82}, are pseudo-magnitudes closely 
related to apparent magnitudes, but minimally affected by uncertainties on reddening.
On the basis of two magnitudes,  $m_{\lambda_1}$ and $m_{\lambda_2}$, we can define a 
Wesenheit index:
\begin{equation}
W(\lambda_2,\lambda_1)=m_{\lambda_1}-\left[\frac{A(\lambda_1)}{E (m_{\lambda_2}-m_{ \lambda_1})}\right]\times (m_{\lambda_2}-m_{ \lambda_1});
\label{eq: W}
\end{equation}
where $ \lambda_1>\lambda_2$ and $\frac{A(\lambda_1)}{E(m_{\lambda_2}-m_{ \lambda_1})}$ 
is the total to selective extinction for the given filters --$\{\lambda_i=V, I, J, H, K_{\rm S}\}$-- 
and for the adopted reddening law. The clear advantage in using the Wesenheit indices is that 
they are minimally affected by uncertainties affecting reddening corrections for Galactic 
and extragalactic Cepheids. Once we fix the reddening law \citep{cardelli89} and we assume 
$R_{V}$=$\frac{A(V)}{A(B)-A(V)}$=3.23, we obtain the following selective absorption ratios, 
namely $A_I$/$A_V$=0.61; $A_J$/$A_V$=0.29; $A_H$/$A_V$=0.18; $A_{K_S}$/$A_V$=0.12 mag.

By combining the five  optical--NIR (\textit{VIJHK$_{\rm S}$})  mean magnitudes, we can compute 
10 Wesenheit indices for each Cepheid in the sample, and in turn, 10 PW 
relations of the form  $W(\lambda_2,\lambda_1)=a+b\times\log P$, where P is the 
pulsation period in days.
We decided to analyze separately FU and FO Cepheids to overcome possible systematic 
uncertainties that the 'fundamentalization' of the FO periods might introduce in the 
estimate of the PL relations \citep{feast97,marengo10}. 
Therefore, we computed independent PW relations for FU and FO Cepheids.
We performed a linear fit of the data to identify possible outliers. We have 
included data up to 4$\sigma$ from the central location by 
using the robust \textsl{Biweight} location estimator \citep{bw}.
We ended up with a sample of $\sim$4000 SMC ($\sim$2500, FU; $\sim$1500, FO) and 
$\sim$3000 LMC ($\sim$1800, FU; $\sim$1100, FO) Cepheids. For approximately three 
dozen Cepheids with period $\gtrsim$ 20 days the $I$ band is saturated in the OGLE-III 
data-set, and therefore we cannot apply the template. The NIR mean magnitudes for 
these Cepheids were taken from P04 and transformed into the 2MASS photometric system 
(see green dots in Figure~2).

We then performed a linear regression of the NIR data and the results for the 
three PW relations are showed in Figure~2 (see also Table~1).
From top to bottom each panel shows FU (red and green dots) and FO (blue dots) 
LMC (left) and SMC (right) Cepheids. The PW relations are over--plotted as black lines.  
Data plotted in Figure~2 display four relevant findings concerning the NIR PW relations. 
(1)The FU and the FO PW relations are linear over the entire period range.
(2)The intrinsic dispersion of the SMC PW relations are a factor of two 
larger than for the LMC PW relations. The difference is mainly caused by depth effects 
in the former system \citep{vandenbergh00}.
(3)For each PW relation the difference in the slope between LMC and 
SMC Cepheids is small. Data listed in Tables~1 and ~2 indicate that 
it is, on average, smaller than 0.8$\sigma_{tot}$, where $\sigma_{tot}$ is 
the sum in quadrature of the dispersion of LMC and SMC PW relations.

This indicates a minimal dependence of the NIR PW relations on the metal content. 
(4)The width in temperature of the FO instability strip is narrower than the instability strip 
for FU Cepheids \citep{bono00}, but the intrinsic dispersions are not significantly different.  
The lack of a template light curve for FO Cepheids increases the dispersion of their mean 
magnitudes.  

The above findings support recent theoretical \citep{bono10} and empirical 
\citep{majess11} investigations. The main advantage of the current approach 
is that the results rely on NIR single epoch measurements that are 2--3 times 
larger than any previous investigation \citep[][]{groenewegen00}.
In order to constrain the possible occurrence of systematic errors in the 
NIR PW relations, we also computed the optical--NIR PW relations using the 
$V$, $I$ mean magnitudes provided by OGLE-III. 
Figures 3 and 4 show the optical--NIR PW relations for FU (red dots) and 
FO (blue dots) LMC and SMC Cepheids, respectively. Once again, we found 
that the PW relations are linear and the slopes are minimally affected 
by the difference in metal content (see Table~1).
Current results concerning the linearity of NIR and optical--NIR PW relations 
support previous findings by \citet{ngeow05} and \citet{madore09}.

\subsection{Linearity of PW Relations} 

To constrain on a more quantitative basis the linearity of NIR PW relations
we estimated the distance of individual Cepheids from the least squared 
solution. 
The residuals  for FU Cepheids plotted in the top panels of Figure~5 do not 
show any trend as a function of the pulsation period. To further constrain this
evidence, we performed a linear fit to the residuals and we found that the 
zero--points, the slopes and the means attain vanishing values. Moreover,
the dispersions are typically smaller than 0.2 mag.
The same outcome applies to the FO Cepheids (see bottom panels of Figure~5). 
Note that the residuals of FO PW relations are larger than the residuals of
the FU PW relations, since for the former ones we lack template light curves.
The residuals referred to SMC are larger than the residuals of LMC due to depth
effects.
The anonymous referee explicitly asked a quantitative analysis concerning
the linearity of the PW relations for both FU and FO SMC Cepheids. To our
knowledge there is no clear physical reason why FU and FO NIR PW relations
should show a break, therefore, we decided to constrain the possible occurrence
using different breaks in period. We split the entire Cepheid sample by adopting
a break in period at $\log P$=0.45. This means that we assume as short-period
Cepheids those with $\log P\le$0.45, while the long--period ones are those
with 0.45$< \log P \le$1.65 (FU) and with 0.45$< \log P\le $0.65 (FO).
The zero--points and the slopes for FU NIR PW relations listed in Table~3
show that their errors are a factor of 3--4 larger than the errors of the linear regressions
based on the entire sample. This trend is expected, since the number of Cepheids
included in the two new linear regressions decreases from a factor of three
(long--period) to 50\% (short--period). On the other hand, the dispersions
of the new PW relations are either similar (short--period) or on average
smaller (long--period). The new FO PW relations show similar trends
concerning the intrinsic errors on the zero--points, on the slopes
and on the dispersions.

The break in period is defined somewhat arbitrarily, therefore, we decided
to perform the same test, but using a break at $\log P$=0.40 and $\log P$=0.35.
Current empirical evidence suggests that optical PL relations of SMC Cepheids 
show a break in period at $\log P$$\approx$0.4 \citep{sandage09}, while for LMC 
Cepheids the break seems to be at $\log P$$\approx$1 \citep{sandage04}.
The results concerning the new NIR PW relations are listed in Table~3 and
indicate that the short--period PW relations are quite similar to the
global PW relations, i.e., the PW relations covering the entire period range.
This trend is --once again-- expected, since more than 2/3 of the Cepheid
sample is in the short--period range.
The evidence that linear regressions with an arbitrary break in period,
give PW relations with either similar or marginally smaller dispersion is also expected.
This is the consequence of the increase in the degrees of freedom of the linear
regressions. However, this does not mean that the PW relations with a break in
period are a better representation of observations. To address this issue on
a more quantitative basis, we devised a new empirical test based on the relative
distance between SMC and LMC. The MC relative distance is quite solid, since
different standard candles provide similar estimates.

By adopting both NIR and optical--NIR PW relations, we found that the
relative distance modulus based on FU Cepheids and at $\log P$=1 is
$\Delta \mu$=0.48$\pm$0.03 mag. This evaluation agrees quite well with
similar estimates available in the literature (see Section~4). To further constrain
the intrinsic accuracy of the NIR PW relations with a break in period, we
computed  three new  PW($J$,$K_{\rm S}$) relations for LMC Cepheids. Following
 \citet{sandage04}, we adopted a break in period
at $\log P$=1. The zero--points and the slopes of the new NIR PW relations
are listed in Table~4. We estimated the MC relative distances by using
the short--period and the long--period PW relations. The relative
distances based on the former ones  were estimated at $\log P$=0.3,
while those based on the latter ones were estimated at $\log P$=1.0.
The results listed in Table~5 indicate --as expected-- that the MC relative
distances based on short--period PW relations agree quite well with the
MC relative distances based on global PW relations.
The main difference is that the relative distance based on short--period
PW relations have intrinsic errors, estimated by propagating the errors 
on both the coefficients and the dispersion of the individual PW relations, 
that are on average a factor of two larger than those ones based on the 
global PW relations. The same outcome applies to the MC relative distances 
based on the long--period PW relations. However, their intrinsic errors are 
larger and they also show a larger spread among the three different NIR
PW relations.
Note that the MC relative distance based on the long--period PW($J$,$H$) relations 
are systematically smaller than the others, because the zero--point of the 
long--period PW relation for LMC Cepheids is larger than the zero--point of the 
global PW relation (15.949 versus 15.876).

We repeated the same test by using two different pivot periods,
namely $\log P$=0.2 for the short--period and $\log P$=1.2 for the long--period
PW relations  and the results are quite similar. We also performed the same test
using NIR FO PW relations and the outcome is --once again-- quite similar.
Note that the intrinsic errors on the coefficients of the long--period
FO PW relations are larger than the errors of the short--period ones, since
the Cepheid sample in the former period interval is from a factor of five to
a factor of 10 smaller than in the latter one.

The above findings indicate that the PW relations with arbitrary breaks in
period when compared with global PW relations have larger intrinsic errors
on the coefficients of the linear regressions and roughly equivalent dispersions.

However, the MC relative distances based on the former ones are characterized
by intrinsic errors that are, on average, a factor of two larger than the
latter ones. Thus further supporting the use of global NIR PW relations.

This provides an independent support to the results concerning the linearity 
of both optical and NIR PW relations for FU Cepheids by 
\citet[][and references therein]{persson04,bono10,ngeow12}. 
We found that optical and NIR PW relations for FO Cepheids
are also linear over the entire period range, 
supporting previous findings by \citet{ngeow05} and \citet{madore09}.
We are thus facing the empirical evidence that optical 
and NIR PL relations for FU Cepheids do show a change in the slope for 
$\log P\approx$0.4 \citep{sasselov97, bono99, ngeow05,koen07,matsunaga11}. 
The difference between the PL and the PW relations 
is mainly due to the fact that the latter is mimicking, as originally 
suggested by \citet{bonomarconi99}, a PLC relation.

\subsection{Metallicity dependence of the PW relations} 

To further constrain the metallicity dependence of the NIR PW relations, 
we performed a detailed comparison with similar estimates available in 
the literature. 
The middle panel of Figure~6 shows the difference between the slope of the 
PW($J$,$K_{\rm S}$) relations we estimated for LMC (black line)
and SMC (green line) Cepheids with similar PW relations for Galactic
Cepheids (see Table~6) derived by \citet[][hereinafter S11a; red line]{storm}
and by \citet[][hereinafter N12; purple line]{ngeow12}.
The standard deviations plotted in the same figure clearly indicate
that current Magellanic and Galactic NIR PW relations do agree within 
1$\sigma$. The difference in the slope between our SMC and Galactic 
PW($J$,$K_{\rm S}$) relations is, on average, smaller than 
0.3$\sigma$ (N12) and 0.4$\sigma$ (S11a).

The anonymous referee suggested to perform the same comparison for the optical 
PW($V$,$I$) relation. The top panel of Figure~6 shows the difference between 
the slope of the PW($V$,$I$) relations we estimated for LMC (black line) and 
SMC (green line) Cepheids with similar PW relations for Galactic Cepheids 
(see Table~6) derived by S11a (red line) and  
\citet[][hereinafter B07; blue line]{benedict07}. 
The standard deviations plotted in the same figure clearly indicate that 
current Magellanic and Galactic optical PW relations do agree within 1$\sigma$. 
The difference in the slope of the PW($V$,$I$) relation between 
our metal-poor stellar system (SMC, [Fe/H]=-0.75) and 
our metal-rich stellar system (Galaxy, [Fe/H]= -0.18 to +0.25) is, 
on average, smaller than $\sim$0.1$\sigma$ (B07) and 
$\sim$0.9$\sigma$ (S11a).
The bottom panel of Figure~6 shows the difference between the slope of the 
PW($V$,$K_{\rm S}$) relations we estimated for LMC (black line) and
SMC (green line) Cepheids with the PW relation for LMC Cepheids 
(see Table~6) derived by R12 (gray line). 
Data plotted in this figure clearly indicate the good agreement between 
the two independent LMC slopes. Moreover, current SMC and LMC PW relations 
do agree within 1$\sigma$.
The other NIR and optical-NIR PW relations provide similar results. 
The quoted numbers indicate that the PW relations are, in the metallicity range 
covered by Magellanic Cepheids, independent of metal abundance. 
The extension into the more metal-rich regime does require more accurate measurements 
for Galactic Cepheids.

\subsection{Absolute calibration of the PW relations} 

To estimate the distances to the MCs, we combined our new comprehensive sets 
of PW relations with recent findings concerning absolute magnitudes of Galactic 
Cepheids. We followed the same approach suggested by P04 to calibrate the 
LMC PW relations and adopted the 10 FU Galactic Cepheids with \textit{Hubble Space 
Telescope} trigonometric parallaxes \citep{benedict07}. 
To calibrate the FO PW relations, we adopted the {\textit Hipparcos} trigonometric 
parallaxes for Polaris provided by \citet{vanleeuwen07}.
The mean \textit{J,H,K$_{\rm S}$} magnitudes for the 
calibrating Galactic Cepheids are from \citet{laney92}. 
We estimated the true distance modulus --$\mu$-- of both LMC and SMC 
by using the quoted calibrators and by imposing the slope of individual 
PW relations for FU and FO Cepheids (see Column 6 of Table~1). 

Note that the true distance moduli for FU Cepheids were estimated as 
the weighted mean of the $\mu_i$ of individual calibrating Cepheids. 
The associated error on $\mu$ is the sum in quadrature of the 
weighted error on the distance and of the intrinsic dispersion 
associated with the linear regression (see Column 5 of Table~1).
The weighted means based on the FU PW relations give  
$\mu$(LMC)=18.45$\pm$0.02 and $\mu$(SMC)=18.93$\pm$0.02 mag, 
while the weighted means based on FO PW relations give 
$\mu$(LMC)=18.60$\pm$0.03 and $\mu$(SMC)=19.12$\pm$0.03 mag.  

To constrain the possible occurrence of deceptive errors in the absolute 
zero-point, we performed an independent zero-point calibration using 
predicted FU PW relations for Magellanic Cepheids provided by 
\citet{bono99} and  \citet{marconi05}. Recent investigations indicate 
that theory and observations agree quite well concerning optical and 
optical-NIR PW relations \citep{bono10}. The true distance moduli 
based on the new zero-point calibration are listed in column seven
of Table~1. The error associated to individual distance moduli is 
the standard deviation from the theoretical PW relation.
The new weighted means based on the FU PW relations give
$\mu$(LMC)=18.56$\pm$0.02 and $\mu$(SMC)=18.93$\pm$0.02 mag.  
Interestingly enough, the two independent calibrations for FU Cepheids 
do provide weighted true distance moduli to the MCs that agree 
quite well (DM $\lesssim 0.11$~mag). This finding appears  
even more compelling if we take into account that we are using 
independent NIR and optical data sets together with independent 
theoretical and empirical calibrators. 

Current zero-point calibration for FO PW relations relies on the 
trigonometric parallax of a single object \citep[Polaris,][]{vanleeuwen07}. 
Absolute distances for FO Galactic and Magellanic Cepheids based on the 
IRSB method are not available. To overcome this problem, we decided to use 
predicted FO PW relations for Magellanic Cepheids provided by 
\citet{bono99} and \citet{marconi05}. The true distance moduli 
based on the new zero-point calibration are listed in Column 7
of Table~1. The error associated to individual distance moduli are  
the dispersions from the theoretical PW relation.
The new weighted means based on the FO PW relations give
$\mu$(LMC)=18.51$\pm$0.02 and $\mu$(SMC)=19.02$\pm$0.02 mag.  
The two independent empirical calibrations for FU and FO Cepheids provide 
weighted true distance moduli to the MCs that differ from 0.15 (LMC) to 
0.19 (SMC) mag. On the other hand, the weighted true distance moduli 
based on the theoretical calibrations differ at the level of a few 
hundredths of mag. The difference between the two empirical 
calibrations is due to the fact that the empirical calibrations for 
FO PW relations rely on a single object (see Section~4). 

However, data listed in Table~1 indicate that the PW($J,H$) and the PW($I,H$) 
relations for FU and FO Cepheids, calibrated using the Galactic Cepheids 
with trigonometric distances, provide true distance moduli that differ 
at the 2$\sigma$-3$\sigma$ level from the weighted mean. The evidence that 
distances based on PW relations, calibrated using theoretical predictions 
for MC Cepheids (L. Inno et al. 2013, in preparation), show smaller 
differences indicates that the main culprit seems to be the precision 
of the $H$ band zero-point calibration. However, the difference in the 
weighted mean distances, provided by the two independent zero-point 
calibrations for FU Cepheids, is smaller than 
5\% (LMC: 49.0 $\pm$1.2 versus 51.5$\pm$1.2 kpc; 
SMC: 61.1$\pm$2.2 versus 68.8$\pm$2.3 kpc). 
Moreover, the total uncertainty of current LMC and SMC distances 
is at the $\sim$2\% and at the $\sim$4\% level, respectively. 
Note that we obtain very similar distances if we neglect the distances based on 
the PW($J,H$) and PW($I,H$) relations, namely 18.47$\pm$0.02 (trigonometric 
parallaxes) and 18.57$\pm$0.03 (theory) mag.

To further constrain the possible sources of systematic errors in current distance 
estimates, we also constrained the impact of the adopted reddening law. In a recent 
investigation \citet{kudritzki11} suggested that distance determinations based on 
the PW relations may be affected by changes in the reddening law either in the 
Galaxy or in external stellar systems. To constrain this effect we computed a new 
set of PW relations by adopting the reddening law by \citet{mccall04}. We found 
that the difference in the true distance moduli, based on the two different 
reddening laws, is on average $\sim$0.01 mag. 
The mild dependence of current distance determinations on the reddening law 
might also be due to the fact that the selective absorption ratios of optical--NIR 
PW relations are less sensitive to the fine structure of the reddening law. 
The selective absorption ratios given in Section~3 indicate that the coefficient of 
the color term of the PW($V,K$) relation is at least one order of magnitude 
smaller than the coefficients of PW($J,H$) and PW($H,K_{\rm S}$) relations  
(0.13 versus 1.63 and 1.92 mag).
This evidence indicates that the difference in the true distance moduli 
based on PW($J,H$) and PW($H,K_{\rm S}$) relations might also be caused 
either by photometric error in the mean magnitudes or by changes in the 
reddening law along the line-of-sight of the \textit{HST} Galactic calibrating 
Cepheids.

\section{Summary and discussion}

We present new true distance modulus determinations of the MCs using 
NIR and optical-NIR PW relations. The NIR PW relations were estimated 
adopting the largest data set of \textit{J,H,K$_{\rm S}$} single epoch measurements 
ever collected for MC Cepheids. The optical $V,I$ measurements come from the 
OGLE-III data set. We ended up with a sample of 4150 (2571, FU; 1579, FO; SMC) 
and 3042 (1840, FU; 1202, FO; LMC) Cepheids.
We estimated independent PW relations for both FU and FO Cepheids. 
The slopes of the current FU PW relations agree quite well with similar 
estimates available in the literature. We found that they agree at 1.2$\sigma$
level with the slopes of the NIR PW relations for LMC Cepheids derived by P04. 
The agreement is even better if we compare our slopes for the PW($J,K_{\rm S}$) 
relations with the slopes recently provided by \citet[][hereinafter S11b]{stormb} 
for the LMC (LMC: -3.31$\pm$0.09 versus -3.365$\pm$0.008). 
The above findings are even more relevant if we take into account that 
current slopes are based on data samples that are from 
80 (S11b) -- 30 (P04) times to $\sim$ 3 times 
\citep{groenewegen00} larger than the quoted samples.
We cannot perform a similar comparison concerning the slopes of the 
FO PW relations, since to our knowledge they are not available in the 
literature.   

Moreover, we found that both FU and FO PW relations are linear over 
the entire period range and their slopes attain, within the intrinsic 
dispersions, similar values in the MCs. The difference is, on average, 
smaller than 0.8$\sigma$. 
The difference between the slope of our SMC and Galactic
PW($J$,$K_{\rm S}$) relations available in the literature
is, on average, smaller than 0.5$\sigma$ (0.3$\sigma$,N12;
0.4$\sigma$, S11b). The same outcome applies to optical bands,
and indeed the difference in the slope between our SMC
and Galactic PW($V$,$I$) relations available in the literature
is, on average, smaller than $\sim$0.1$\sigma$ (B07) and
$\sim$0.9$\sigma$ (S11a). This supports the evidence for
a marginal dependence of NIR and PW(V,I) relations 
on the metal content, as suggested by pulsation predictions 
and recent empirical investigations.

The new PW relations were calibrated using two independent sets of 
Galactic Cepheids with individual distances based either on 
trigonometric parallaxes or on theoretical models.
By using FU Cepheids we found a true distance modulus to the LMC of
18.45$\pm$0.02 (random) $\pm$0.10 (systematic) mag and to the SMC of
18.93$\pm$0.02 (random) $\pm$0.10 (systematic) mag.  
These estimates are the weighted mean over the entire set of distance 
determinations. The random error was estimated by taking into account
the intrinsic dispersion of individual PW relations. The systematic 
error is the sum in quadrature of the difference in $\mu$ introduced 
by the change in reddening law and in the zero-point 
calibration.

We found similar distances using FO Cepheids 
18.60$\pm$0.03 (random) $\pm$0.10 (systematic) mag, LMC 
and 19.12$\pm$0.03 (random) $\pm$0.10 (systematic) mag, SMC.  
Once again the random errors were estimated by taking into
account the intrinsic dispersion of individual PW relations,
while the systematic ones are the sum in quadrature of the
difference in $\mu$ introduced by the change in reddening law
and in the zero-point calibration. 

The two independent empirical calibrations for FU and FO Cepheids provide
weighted true distance moduli to the MCs that differ for 0.15 (LMC) and 
0.19 (SMC) mag. On the other hand, the weighted true distance moduli
based on the theoretical calibrations differ at the level of a few
hundredths of mag. The difference between the empirical and theoretical
calibrations is due to the fact that the empirical calibrations for
FO PW relations rely on a single object (see Section~3.2).

The relative distance of the MCs, for distance indicators minimally 
affected by the metal content, is independent of uncertainties affecting 
the zero-point calibration. We found that the weighted mean relative distance 
between SMC and LMC using FU Cepheids and the PW relation listed in Table~1 
($\log P$=1) is $\Delta\mu$=0.48$\pm$0.03 mag. We applied the same 
approach by using FO Cepheids and we found $\Delta\mu$=0.52$\pm$0.03 mag 
($\log P$=0.5). The errors on the weighted mean relative distances were 
estimated by using the dispersions of individual PW relations. 
The quoted determinations agree quite well with each other and with the 
recent estimate $\Delta\mu$=0.47$\pm$0.15 mag provided by S11b  
by using the IRSB method \citep[see also][]{groenewegen00,bono10,matsunaga11}.

The distance modulus we obtained for the LMC agrees quite well with the recent 
estimate provided by S11b (18.45$\pm$0.04 mag) and by P04 
(18.50$\pm$ 0.05mag) by using the NIR PL, 
PLC and PW relations. The difference is also minimal 
with the ``classical" value --18.50$\pm$0.10 mag-- \citep{freedman01}.   
The same conclusion can be reached if we compare the current estimate 
with recent distance moduli provided by \citet[][18.50$\pm$0.03; 
$HST$ trigonometric parallaxes for Galactic Cepheids and the LMC slope of 
the optical PW relation]{benedict07};
by \citet[][18.49$\pm$0.04; optical PL and PLC relations]{ngeow08}; 
by \citet[][18.44$\pm$0.03 (random) $\pm$0.06 (systematic); PW($V,I$) relation for Galactic and 
LMC Cepheids ]{madore10}
and by \citet[][18.531$\pm$0.043 mag; NIR and optical-NIR 
PW relations]{ngeow12} \footnote{Note that in the comparison of LMC distance moduli, 
we adopted the estimates that neglect the metallicity dependence.}.
Moreover, our result also agrees with the most recent distance 
modulus -- 18.477 $\pm$ 0.033-- provided by \citet{scowcroft11} and \citet{freedman12}, 
using the $Spitzer$  mid-IR band PL relations.

The distance modulus we obtained for the SMC is, once again, in very good 
agreement with the independent estimates provided by \cite[][19.11$\pm$0.11 mag; \textit{Hipparcos} trigonometric parallaxes and PW($V,I$) relation]{groenewegen00} 
 and  S11b (18.92$\pm$0.14 mag).

\section{Final remarks}

The key feature of current findings is that the random errors associated to 
our distance determinations are very small, due to the fact that we adopt an 
homogeneous and accurate NIR data set and also because we are fully exploiting 
the use of NIR and optical--NIR PW relations.    
Moreover, the use of two independent zero-point calibrations and two different 
reddening laws indicate that the global uncertainty on the MC distances 
seems to be of the order of 1\% by using either the 10 NIR/optical--NIR 
PW relations or the seven optical--NIR PW relations.

However, there are a few pending issues that need to be addressed in more 
detail in the near future. 
 
\begin{enumerate}

\item The very good intrinsic accuracy of NIR and optical--NIR PW relations further 
support the findings by \citet[][see their Figures 13 and 14]{bono10} indicating 
that the difference between optical ($B,V$) and NIR ($J,K_{\rm S}$) PW relations 
can be adopted to constrain the metallicity correction(s) to the Cepheid 
distance scale based on optical PL relations.
Moreover, current findings indicate that the error 
budget of the absolute distances based on PW relations is dominated 
by uncertainties in the zero--point. The solution of this problem 
appears quite promising in light of the fact that \textit{Gaia} will be 
launched in approximately one year and the number of double eclipsing 
binaries including classical Cepheids is steadily increasing during 
the last few years \citep{pietrzynski10,pietrzynski11}. Moreover, new 
optical  \citep[OGLEIV][]{sos2012} and NIR  
\citep[Galaxy:VVV,][]{minniti10}; \citep[MCs:VMC,][R12]{cioni11} 
surveys will also provide new, homogenous and accurate mean magnitudes.

\item The above results provide an independent support to the 
plausibility of the physical assumptions adopted in current hydrodynamical
models of variable stars. Indeed the distance moduli based on theoretical 
calibrations  agree well with distance moduli based on empirical calibration. 
However, we still lack detailed investigations concerning the pulsation 
properties of Classical Cepheids in the metal-intermediate regime. 
In particular, we need a comprehensive analysis of the metallicity dependence 
of both PW relations and Period-Luminosity-Color relations in the optical 
and in the NIR regime.    
 
\item Accurate spectroscopic iron abundances are only available 
for roughly the 50\% of Galactic Cepheids \citep[][and references therein]{romaniello08,pedicelli09,luck11}
and for a few dozen of MC Cepheids. However, the empirical scenario 
will have a relevant jump thanks to the ongoing massive ground-based 
spectroscopic surveys at the 8m class telescopes \citep[\textit{Gaia ESO Survey},][]{gilmore12,tolstoy09}

\item  Plain physical arguments indicate that FO Cepheids have 
the potential to be robust distance indicators \citep{bono00}. However,  
we still lack for these variables NIR template light curves. Moreover, 
current FO absolute calibrations are also hampered by 
the lack of precise distance determinations based on trigonometric 
parallaxes for a good sample of Galactic calibrators. These circumstantial 
evidence limits the precision of MC distance determinations based on 
FO Cepheids. 

\item  Absolute distances based on PW relations including the 
H band are characterized by a large spread. The reasons for this 
behavior are not clear. No doubt that new high-resolution, high 
signal-to-noise NIR spectra of Galactic Cepheids \citep{bono12}
can shed new lights on this open problem.    

\end{enumerate}

\acknowledgments
It is a pleasure to thank an anonymous referee for his/her pertinent 
suggestions and criticisms that helped us to improve the readability of 
the paper. We also acknowledge M. Fabrizio for many useful discussions 
concerning the use of Biweight and data set cleaning.  
One of us (G.B.) thanks the ESO for support as a science visitor.
This work was partially supported by PRIN-INAF 2011 ``Tracing the
formation and evolution of the Galactic halo with VST" (P.I.: M. Marconi)
and by PRIN-MIUR (2010LY5N2T) ``Chemical and dynamical evolution of
the Milky Way and Local Group galaxies" (P.I.: F. Matteucci).


%
\clearpage

\begin{deluxetable}{llrrcll}
\tabletypesize{\scriptsize}
\tablewidth{0pt}

\tablecaption{NIR and Optical-NIR PW relations for LMC and SMC Cepheids.}
\label{tab}
	
\tablehead{
\colhead{W($\lambda_2$,$\lambda_1$) \tablenotemark{a}}&
\colhead{Mode}&
\colhead{a}&
\colhead{b}&
\colhead{$\sigma$\tablenotemark{b}}&
\colhead{$\mu_{\pi}$}&
\colhead{$\mu_{theo}$}
}
\startdata
\multicolumn{7}{c}{LMC} \\
\\

W($J,K_{\rm S}$)  & FU (1708)&  15.876  $\pm$ 0.005 &   -3.365  $\pm$ 0.008&0.08  & 18.44 $\pm$0.05\tablenotemark{c} & 18.53 $\pm$ 0.07\tablenotemark{d}\\
W($J,H$)          & FU (1701)&  15.630  $\pm$ 0.006  & -3.373   $\pm$ 0.008& 0.08 & 18.30 $\pm$0.05\tablenotemark{c} & 18.65 $\pm$ 0.04\tablenotemark{d}\\
W($H,K_{\rm S}$)  & FU (1709)&  16.058  $\pm$ 0.006  & -3.360   $\pm$ 0.010& 0.10 & 18.54 $\pm$0.05\tablenotemark{c} & 18.46 $\pm$ 0.12\tablenotemark{d}\\
W($V,K_{\rm S}$)  & FU (1737)&  15.901  $\pm$ 0.005  & -3.326   $\pm$ 0.008&0.07  & 18.46 $\pm$0.05\tablenotemark{c} & 18.51 $\pm$ 0.08\tablenotemark{d}\\
W($V,H$)          & FU (1730)&  15.816  $\pm$ 0.005  & -3.315   $\pm$ 0.008&0.07  & 18.40 $\pm$0.05\tablenotemark{c} & 18.56 $\pm$ 0.06\tablenotemark{d}\\
W($V,J$)          & FU (1732)&  15.978  $\pm$ 0.006  & -3.272   $\pm$ 0.009&0.08  & 18.49 $\pm$0.05\tablenotemark{c} & 18.47 $\pm$ 0.12\tablenotemark{d}\\
W($I,K_{\rm S}$)  & FU (1737)&  15.902  $\pm$ 0.005  & -3.325   $\pm$ 0.008&0.07  & 18.46 $\pm$0.05\tablenotemark{c} & 18.52 $\pm$ 0.08\tablenotemark{d}\\
W($I,H$)          & FU (1734)&  15.801  $\pm$ 0.005  & -3.317   $\pm$ 0.008&0.08  & 18.39 $\pm$0.05\tablenotemark{c} & 18.55 $\pm$ 0.06\tablenotemark{d}\\
W($I,J$)          & FU (1735)&  16.002  $\pm$ 0.007  & -3.243   $\pm$ 0.011&0.10  & 18.50 $\pm$0.05\tablenotemark{c} & 18.46 $\pm$ 0.12\tablenotemark{d}\\
W($V,I$)          & FU (1700)&  15.899  $\pm$ 0.005  & -3.327   $\pm$ 0.008&0.07  & 18.47 $\pm$0.05\tablenotemark{c} & 18.53 $\pm$ 0.13\tablenotemark{d}\\

&&&&&&\\

MEAN (FU) &&&&& 18.45 $\pm$ 0.02\tablenotemark{e} &    18.56 $\pm$ 0.02\tablenotemark{e}\\
&&&&&&\\

W($J,K_{\rm S}$)& FO  (1057)  & 15.370 $\pm$ 0.005 &   -3.471    $\pm$ 0.013 & 0.08 & 18.60 $\pm$0.08\tablenotemark{f} & 18.52 $\pm$ 0.06\tablenotemark{g}\\
W($J,H$)& FO  (1064)  & 15.207 $\pm$ 0.005 &   -3.507     $\pm$ 0.015&   0.09 &  18.60      $\pm$0.08\tablenotemark{f} & 18.56 $\pm$ 0.06\tablenotemark{g}\\
W($H,K_{\rm S}$)& FO (1063)   & 15.483 $\pm$ 0.007 &   -3.425     $\pm$ 0.017 &0.10 & 18.59 $\pm$0.08\tablenotemark{f} & 18.49 $\pm$ 0.07\tablenotemark{g}\\
W($V,K_{\rm S}$)& FO  (1061) & 15.410 $\pm$ 0.005 &    -3.456    $\pm$ 0.013   &0.07 & 18.61 $\pm$0.08\tablenotemark{f} & 18.51 $\pm$ 0.06\tablenotemark{g}\\
W($V,H$)& FO (1071)  & 15.357 $\pm$ 0.004 &  -3.485    $\pm$ 0.011  &0.08 & 18.61           $\pm$0.08\tablenotemark{f} & 18.52 $\pm$ 0.06\tablenotemark{g}\\
W($V,J$)& FO   (1086) & 15.475 $\pm$ 0.005 & -3.434 $\pm$ 0.014   &0.10 & 18.62           $\pm$0.08\tablenotemark{f} & 18.48 $\pm$ 0.06\tablenotemark{g}\\
W($I,K_{\rm S}$)& FO (1059)  & 15.402  $\pm$ 0.005&  -3.448   $\pm$ 0.013   &0.08 & 18.61   $\pm$0.08\tablenotemark{f} & 18.50 $\pm$ 0.06\tablenotemark{g}\\ 
W($I,H$)& FO (1072)  & 15.351 $\pm$ 0.004 & -3.489  $\pm$ 0.012   &0.08 & 18.62             $\pm$0.08\tablenotemark{f} & 18.51 $\pm$ 0.06\tablenotemark{g}\\
W($I,J$)& FO (1100)   & 15.499 $\pm$ 0.006& -3.423    $\pm$ 0.020  & 0.13 & 18.66           $\pm$0.08\tablenotemark{f} & 18.45 $\pm$ 0.06\tablenotemark{g}\\
W($V,I$)& FO (1081)   & 15.399 $\pm$ 0.003 &-3.460     $\pm$ 0.009  & 0.07 & 18.52          $\pm$0.06\tablenotemark{f} & 18.56 $\pm$ 0.06\tablenotemark{g}\\

\\
 MEAN (FO) &&&&&  18.60$\pm$0.03\tablenotemark{e} &    18.51 $\pm$0.02 \tablenotemark{e}\\

\hline
\multicolumn{7}{c}{SMC} \\
\\
W($J,K_{\rm S}$)  & FU (2448)& 16.457  $\pm$ 0.006 & -3.480  $\pm$ 0.011   &0.16  & 18.92 $\pm$0.05\tablenotemark{c} & 19.01 $ \pm$ 0.10\tablenotemark{d}\\
W($J,H$)          & FU (2448)& 16.217  $\pm$ 0.006 & -3.542  $\pm$ 0.011   &0.17  & 18.74 $\pm$0.05\tablenotemark{c} & 19.02 $ \pm$ 0.07\tablenotemark{d}\\
W($H,K_{\rm S}$)  & FU (2448)& 16.638  $\pm$ 0.006 & -3.445  $\pm$ 0.011   &0.19  & 19.05 $\pm$0.05\tablenotemark{c} & 19.01 $ \pm$ 0.14\tablenotemark{d}\\
W($V,K_{\rm S}$)  & FU (2295)& 16.507  $\pm$ 0.005 &  -3.461 $\pm$ 0.011   &0.15  & 18.95 $\pm$0.05\tablenotemark{c} & 19.00 $ \pm$ 0.11\tablenotemark{d}\\
W($V,H$)          & FU (2285)& 16.426  $\pm$ 0.005 & -3.475  $\pm$ 0.010   &0.15  & 18.88 $\pm$0.05\tablenotemark{c} & 19.00 $ \pm$ 0.10\tablenotemark{d}\\
W($V,J$)          & FU (2286)& 16.614  $\pm$ 0.005 & -3.427  $\pm$ 0.011   &0.16  & 19.00 $\pm$0.05\tablenotemark{c} & 18.98 $ \pm$ 0.14\tablenotemark{d}\\
W($I,K_{\rm S}$)  & FU (2294)& 16.511  $\pm$ 0.005 &  -3.464 $\pm$ 0.011   &0.16  & 18.95 $\pm$0.05\tablenotemark{c} & 19.00 $ \pm$ 0.10\tablenotemark{d}\\
W($I,H$)          & FU (2202)& 16.417  $\pm$ 0.005 & -3.480  $\pm$ 0.011   &0.15  & 18.87 $\pm$0.05\tablenotemark{c} & 19.00 $ \pm$ 0.10\tablenotemark{d}\\
W($I,J$)          & FU (2279)& 16.662  $\pm$ 0.006 &  -3.424 $\pm$ 0.013   &0.18  & 19.01 $\pm$0.05\tablenotemark{c} & 18.92 $ \pm$ 0.14\tablenotemark{d}\\
W($V,I$)          & FU (2260)& 16.482  $\pm$ 0.005 &  -3.449 $\pm$ 0.010   &0.13  & 18.95 $\pm$0.05\tablenotemark{c} & 19.03 $ \pm$ 0.12\tablenotemark{d}\\
\\
MEAN (FU) &&&&& 18.93$\pm$0.02\tablenotemark{e}  &  18.99$\pm$0.03\tablenotemark{e}\\	    

&&&&&&\\
		    
W($J,K_{\rm S}$)& FO  (1461)  & 15.947 $\pm$ 0.005 &-3.651   $\pm$ 0.022 &0.16 & 19.06 $\pm$0.08\tablenotemark{f} & 19.02 $ \pm$   0.04\tablenotemark{g}\\
W($J,H$)& FO  (1473)  & 15.778 $\pm$ 0.006         &-3.722   $\pm$ 0.023 &0.17 & 19.17 $\pm$0.08\tablenotemark{f} & 19.05 $ \pm$   0.03\tablenotemark{g}\\
W($H,K_{\rm S}$)& FO (1456)   & 16.069 $\pm$ 0.007 &-3.579   $\pm$ 0.027 &0.19 & 19.00 $\pm$0.08\tablenotemark{f} & 19.01 $ \pm$   0.04\tablenotemark{g}\\
W($V,K_{\rm S}$)& FO  (1472) & 15.992 $\pm$ 0.005  &-3.624   $\pm$ 0.021 &0.16 & 19.09 $\pm$0.08\tablenotemark{f} & 19.02 $ \pm$   0.04\tablenotemark{g}\\
W($V,H$)& FO (1482)  & 15.937 $\pm$ 0.005          &-3.660   $\pm$ 0.020 &0.15 & 19.16 $\pm$0.08\tablenotemark{f} & 19.03 $ \pm$   0.05\tablenotemark{g}\\
W($V,J$)& FO   (1494)& 16.074 $\pm$ 0.006          &-3.578   $\pm$ 0.023 &0.18 & 19.17 $\pm$0.08\tablenotemark{f} & 19.02 $ \pm$   0.04\tablenotemark{g}\\
W($I,K_{\rm S}$)& FO (1471)  & 15.990  $\pm$ 0.005 &-3.630   $\pm$ 0.020 &0.16 & 19.09 $\pm$0.08\tablenotemark{f} & 19.02 $ \pm$   0.05\tablenotemark{g}\\
W($I,H$)& FO (1477)  & 15.932 $\pm$ 0.005          &-3.667   $\pm$ 0.020 &0.16 & 19.17 $\pm$0.08\tablenotemark{f} & 19.02 $ \pm$   0.04\tablenotemark{g}\\
W($I,J$)& FO (1499)   & 16.113 $\pm$ 0.007         &-3.595   $\pm$ 0.027 &0.20 & 19.17 $\pm$0.08\tablenotemark{f} & 18.00 $ \pm$   0.05\tablenotemark{g}\\
W($V,I$)& FO (1465)   & 15.958 $\pm$ 0.005         &-3.599   $\pm$ 0.019 &0.14 & 19.12 $\pm$0.06\tablenotemark{f} & 19.05 $ \pm$   0.03\tablenotemark{g}\\

\\
MEAN (FO) &&&&&  19.12$ \pm$ 0.03\tablenotemark{e} &  19.02 $\pm$0.02\tablenotemark{e}\\	    
\enddata    

\tablenotetext{a}{The color coefficients of the adopted PW relations are the following: 
$\frac{A_K}{ E(J-K_{\rm S})}$=0.69; 
$\frac{A_H}{E(J-H)}$=1.63; 
$\frac{A_K}{E(H-K_{\rm S})}$=1.92; 
$\frac{A_K}{E(V-K_{\rm S})}$=0.13; 
$\frac{A_H}{E(V-H)}$=0.22; 
$\frac{A_J}{E(V-J)}$=0.41; 
$\frac{A_K}{E(I-K_{\rm S})}$=0.24; 
$\frac{A_H}{E(I-H)}$=0.42; 
$\frac{A_J}{E(I-J)}$=0.92; 
$\frac{A_I}{E(I-V)}$=1.55}  
\tablenotetext{b}{Dispersion of the linear fit (mag)}
\tablenotetext{c}{Distance modulus based on the zero-point calibration from \citet{benedict07}.}
\tablenotetext{d}{Distance modulus based on the  zero-point calibration obtained by the predicted 
FU PW relations for Magellanic Cepheids provided by  \citet{marconi05}. The associated error is the 
dispersion of the theoretical PW relation.}
\tablenotetext{e}{Weighted distance modulus estimated using the distance moduli of individual PW relations.}
\tablenotetext{f}{Distance modulus obtained using Polaris for the zero-point calibration \citep{vanleeuwen07}.}
\tablenotetext{g}{Distance modulus based on the  zero-point calibration obtained by the predicted 
FO PW relations for Magellanic Cepheids provided by  \citet{marconi05}. The associated error is the 
dispersion of the theoretical PW relation.}

\end{deluxetable} 


\begin{deluxetable}{llrrcll}
\tabletypesize{\scriptsize}
\tablewidth{0pt}

\tablecaption{Difference in the slopes of the PW relations for LMC and SMC Cepheids.
}
\label{tab}
	
\tablehead{
\colhead{W($\lambda_2$,$\lambda_1$)}&
\colhead{Mode}&
\colhead{$\Delta$b$_{LMC-SMC}$\tablenotemark{a}}&
\colhead{$\sigma_{tot}$\tablenotemark{b}}
\\
}
\startdata
W($J,K_{\rm S}$)& FU &      0.115  $\pm$ 0.014   & 0.18 \\
W($J,H$)& FU  &   0.169    $\pm$ 0.014 & 0.19\\
W($H,K_{\rm S}$)& FU  &   0.120    $\pm$ 0.015 & 0.21 \\
W($V,K_{\rm S}$)& FU &    0.135    $\pm$ 0.014   & 0.17 \\
W($V,H$)& FU &   0.160   $\pm$ 0.013   &0.17 \\
W($V,J$)& FU &     0.155   $\pm$ 0.014   &0.18 \\
W($I,K_{\rm S}$)& FU &   0.139    $\pm$ 0.014   & 0.17 \\
W($I,H$)& FU &   0.163    $\pm$ 0.014   & 0.18 \\
W($I,J$)& FU &    0.181 $\pm$ 0.017   & 0.21  \\
W($V,I$)& FU &   0.122   $\pm$ 0.014   & 0.15  \\

&&&\\

\\
W($J,K_{\rm S}$)& FO   &  0.180    $\pm$ 0.026 & 0.18 \\
W($J,H$)& FO   & 0.215	    $\pm$ 0.027	&   0.19 \\
W($H,K_{\rm S}$)& FO   & 0.154      $\pm$ 0.032    &0.21\\
W($V,K_{\rm S}$)& FO   &   0.172    $\pm$ 0.024 &0.18 \\
W($V,H$)& FO  &   0.175     $\pm$ 0.023  & 0.17 \\
W($V,J$)& FO  &   0.143   $\pm$ 0.027   &0.21 \\
W($I,K_{\rm S}$)& FO  & 0.182    $\pm$ 0.024   &0.18 \\ 
W($I,H$)& FO &  0.178      $\pm$ 0.023   &0.18 \\
W($I,J$)& FO    &   0.172     $\pm$ 0.034  & 0.23 \\
W($V,I$)& FO    &    0.139        $\pm$ 0.021  & 0.16 \\
		    
\enddata

\tablenotetext{a}{The error on the difference was estimated by accounting for the uncertainties 
on the slopes of PW relations.}   
\tablenotetext{b}{Total dispersion, i.e. $\sigma_{tot}$=$\sqrt{\sigma_{LMC}^2+\sigma_{SMC}^2}$, 
where $\sigma_{LMC}$ and $\sigma_{SMC}$ are the individual dispersions of the PW relations for 
LMC and SMC Cepheids, respectively.}

\end{deluxetable}

													  
\begin{deluxetable}{llrrrlr}
\tabletypesize{\scriptsize}
\tablewidth{0pt}

\tablecaption{NIR and optical--NIR PW relations for SMC Cepheids computed by assuming  
different break points in period. 
}
\label{tab}
	
\tablehead{
\colhead{W($\lambda_2$,$\lambda_1$)}&
\colhead{Mode}&
\colhead{Period Range}&
\colhead{a}&
\colhead{b}&
\colhead{$\sigma$\tablenotemark{a}}&
\colhead{$N$\tablenotemark{b}}
}
\startdata
\multicolumn{7}{c}{Break point at $\log$ P = 0.35} \\
&&&&&&\\
W($J,K_{\rm S}$)& FU &  $\log$ P$\leqslant$0.35 &  16.532 $\pm$ 0.011&-3.751 $\pm$ 0.052&0.16&        1335\\
W($J,H$)& FU   &   $\log$ P$\leqslant$0.35 & 16.282 $\pm$ 0.011&-3.767 $\pm$ 0.053&0.16&        1343\\
W($H,K_{\rm S}$)& FU  &  $\log$ P$\leqslant$0.35 & 16.712 $\pm$ 0.014&-3.720 $\pm$ 0.066&0.20&        1331\\
W($J,K_{\rm S}$)& FU & 0.35$<$ $\log$ P$\leqslant$1.65 & 16.395 $\pm$ 0.016&-3.429 $\pm$ 0.026&0.14&         942\\
W($J,H$)& FU   &  0.35$<$ $\log$ P$\leqslant$1.65 & 16.156 $\pm$ 0.016&-3.490 $\pm$ 0.026&0.14&         938\\
W($H,K_{\rm S}$)& FU  & 0.35$<$ $\log$ P$\leqslant$1.65 & 16.574 $\pm$ 0.018&-3.392 $\pm$ 0.029&0.16&         936\\
W($J,K_{\rm S}$)& FO & $\log$ P$\leqslant$0.35 & 15.953 $\pm$ 0.006&-3.714 $\pm$ 0.032&0.17&        1206\\
W($J,H$)& FO   &   $\log$ P$\leqslant$0.35 & 15.780 $\pm$ 0.005&-3.757 $\pm$ 0.032&0.17&        1213\\
W($H,K_{\rm S}$)& FO  &  $\log$ P$\leqslant$0.35 & 16.071 $\pm$ 0.007&-3.625 $\pm$ 0.041&0.21&        1204\\
W($J,K_{\rm S}$) & FO & 0.35$<$ $\log$ P$\leqslant$0.65 & 15.870 $\pm$ 0.055&-3.460 $\pm$ 0.120&0.14&         261\\
W($J,H$)         & FO & 0.35$<$ $\log$ P$\leqslant$0.65 & 15.697 $\pm$ 0.053&-3.528 $\pm$ 0.116&0.13&         259\\
W($H,K_{\rm S}$) & FO & 0.35$<$ $\log$ P$\leqslant$0.65 & 15.996 $\pm$ 0.061&-3.405 $\pm$ 0.132&0.15&       260\\
\\
\multicolumn{7}{c}{Break point at $\log$ P = 0.40} \\
&&&&&&\\
W($J,K_{\rm S}$)& FU & $\log$ P$\leqslant$0.40 &  16.531 $\pm$ 0.010&-3.746 $\pm$ 0.043&0.16&        1460\\
W($J,H$)& FU   &   $\log$ P$\leqslant$0.40 & 16.281 $\pm$ 0.010&-3.770 $\pm$ 0.044&0.16&        1464\\
W($H,K_{\rm S}$)& FU  &  $\log$ P$\leqslant$0.40 & 16.710 $\pm$ 0.013&-3.706 $\pm$ 0.056&0.20&        1457\\
W($J,K_{\rm S}$)& FU & 0.40$<$ $\log$ P$\leqslant$1.65 &16.376 $\pm$ 0.019 & -3.403 $\pm$ 0.029 & 0.14 &         816\\
W($J,H$)& FU   &  0.40$<$ $\log$ P$\leqslant$1.65 &  16.137 $\pm$ 0.019 & -3.464 $\pm$ 0.029 & 0.14 &         816\\
W($H,K_{\rm S}$)& FU  & 0.40$<$ $\log$ P$\leqslant$1.65 & 16.543 $\pm$ 0.021 & -3.350 $\pm$ 0.032 & 0.15&         814\\
W($J,K_{\rm S}$)& FO & $\log$ P$\leqslant$0.40 &  15.950   $\pm$ 0.005 &   -3.687    $\pm$ 0.029   & 0.16  & 1277\\
W($J,H$)& FO   &   $\log$ P$\leqslant$0.40 & 15.780 $\pm$ 0.005&-3.748 $\pm$ 0.029&0.16 &        1290\\
W($H,K_{\rm S}$)& FO  &  $\log$ P$\leqslant$0.40 & 16.069 $\pm$ 0.007&-3.599 $\pm$ 0.036 & 0.21&        1278\\
W($J,K_{\rm S}$)& FO & 0.40$<$ $\log$ P$\leqslant$0.65 &  15.779 $\pm$ 0.089&-3.282 $\pm$ 0.181 & 0.14&         186\\
W($J,H$)& FO   &  0.40$<$ $\log$ P$\leqslant$0.65 & 15.570 $\pm$ 0.082&-3.277 $\pm$ 0.167 & 0.13 &         182\\
W($H,K_{\rm S}$)& FO  & 0.40$<$ $\log$ P$\leqslant$0.65 &15.913 $\pm$ 0.096&-3.244 $\pm$ 0.196 & 0.15&         185\\
 \\
\multicolumn{7}{c}{Break point at $\log$ P = 0.45} \\
&&&&&&\\
W($J,K_{\rm S}$)& FU & $\log$ P$\leqslant$0.45 &  16.533 $\pm$ 0.009 & -3.758 $\pm$ 0.038 & 0.16 &        1565\\
W($J,H$)& FU   &   $\log$ P$\leqslant$0.45 & 16.284 $\pm$ 0.010 & -3.787 $\pm$ 0.038 & 0.16 &        1568\\
W($H,K_{\rm S}$)& FU  &  $\log$ P$\leqslant$0.45 & 16.714 $\pm$ 0.012&-3.722 $\pm$ 0.048 & 0.20 &        1563\\
W($J,K_{\rm S}$)& FU & 0.45$<$ $\log$ P$\leqslant$1.65 &  16.375 $\pm$ 0.021&-3.401 $\pm$ 0.031& 0.13 &         707\\\
W($J,H$)& FU   &  0.45$<$ $\log$ P$\leqslant$1.65 & 16.138 $\pm$ 0.022&-3.465 $\pm$ 0.032 & 0.14 &         712\\
W($H,K_{\rm S}$)& FU  & 0.45$<$ $\log$ P$\leqslant$1.65 &16.536 $\pm$ 0.024&-3.339 $\pm$ 0.035 & 0.15 &         711\\
W($J,K_{\rm S}$)& FO & $\log$ P$\leqslant$0.45 & 15.950 $\pm$ 0.005&-3.688 $\pm$ 0.026&0.16&        1335\\
W($J,H$)& FO   &   $\log$ P$\leqslant$0.45 & 15.780 $\pm$ 0.005&-3.754 $\pm$ 0.026&0.16&        1348\\
W($H,K_{\rm S}$)& FO  &  $\log$ P$\leqslant$0.45 &16.069 $\pm$ 0.007&-3.600 $\pm$ 0.033&0.20&        1333\\
W($J,K_{\rm S}$)& FO & 0.45$<$ $\log$ P$\leqslant$0.65 & 15.831 $\pm$ 0.135&-3.379 $\pm$ 0.261&0.14&         128\\
W($J,H$)& FO   &  0.45$<$ $\log$ P$\leqslant$0.65 & 15.613 $\pm$ 0.135&-3.369 $\pm$ 0.262&0.14&         128\\
W($H,K_{\rm S}$)& FO  & 0.45$<$ $\log$ P$\leqslant$0.65 & 15.954 $\pm$ 0.144&-3.321 $\pm$ 0.279&0.15&         127\\
\\

\enddata    
\tablenotetext{a}{Dispersion of the linear fit (mag).}
\tablenotetext{b}{Number of Cepheids adopted in the fit.}

\end{deluxetable} 
					

\begin{deluxetable}{llrrrlr}
\tabletypesize{\scriptsize}
\tablewidth{0pt}

\tablecaption{NIR PW relations for LMC Cepheids computed by assuming a break in period at $\log$P=1}
\label{tab}
	
\tablehead{
\colhead{W($\lambda_2$,$\lambda_1$)}&
\colhead{Mode}&
\colhead{Period Range}&
\colhead{a}&
\colhead{b}&
\colhead{$\sigma$\tablenotemark{a}}&
\colhead{$N$\tablenotemark{b}}
}
\startdata
W($J,K_{\rm S}$)& FU & $\log$P$\leqslant$1.0 & 15.884 $\pm$ 0.007 &-3.380 $\pm$ 0.011 &0.07&        1674\\
W($J,H$)        & FU & $\log$P$\leqslant$1.0 & 15.676 $\pm$ 0.007 &-3.457 $\pm$ 0.012 &0.08 &        1675\\
W($H,K_{\rm S}$)& FU & $\log$P$\leqslant$1.0 & 16.039 $\pm$ 0.008 &-3.324 $\pm$ 0.014 &0.10 &        1684\\
W($J,K_{\rm S}$)& FU & 1.0$<$ $\log$ P$\leqslant$1.65 & 15.950 $\pm$ 0.071 &-3.413 $\pm$ 0.056 &0.08&         69\\
W($J,H$)& FU   &  1.0$<$ $\log$ P$\leqslant$1.65 & 15.778 $\pm$ 0.084 &-3.419 $\pm$ 0.067 & 0.10&         68\\
W($H,K_{\rm S}$)& FU  & 1.0$<$ $\log$ P$\leqslant$1.65 & 16.107 $\pm$ 0.075 &-3.437 $\pm$ 0.060 &0.09&         69\\

\enddata    
\tablenotetext{a}{Dispersion of the linear fit (mag).}
\tablenotetext{b}{Number of Cepheids adopted in the fit.}

\end{deluxetable} 		
										  

\begin{deluxetable}{clrrlll}
\tabletypesize{\scriptsize}
\tablewidth{0pt}

\tablecaption{Difference in distance modulus between LMC and SMC.}
\label{tab}
	
\tablehead{
\colhead{W($\lambda_2$,$\lambda_1$)}&
\colhead{Mode}&
\colhead{$\Delta\mu_{short}$}&
\colhead{$\Delta\mu_{long}$}&
\colhead{Break}&
\colhead{x$_s$\tablenotemark{a}}&
\colhead{x$_l$\tablenotemark{a}}
}
\startdata
W($J,K_{\rm S}$) & FU  &    0.55 $\pm$ 0.06\tablenotemark{b}  & 0.47 $\pm$ 0.06\tablenotemark{b} & \ldots& 0.3 & 1.0 \\
  ''             & FU  &    0.53 $\pm$ 0.11   & 0.48 $\pm$ 0.13  & 0.35 & 0.3 &1.0\\
  ''             & FU  &    0.54 $\pm$ 0.12   & 0.44 $\pm$ 0.13  & 0.40 & 0.3 & 1.0 \\
  ''             & FU  &    0.54 $\pm$ 0.09   & 0.48 $\pm$ 0.13  & 0.45 & 0.3 &1.0\\
\\
   W($J,H$) & FU  &    0.55 $\pm$ 0.05\tablenotemark{b}  & 0.42 $\pm$ 0.06\tablenotemark{b}  & \ldots& 0.3 & 1.0 \\
  '' & FU  &    0.51 $\pm$ 0.11   & 0.31 $\pm$ 0.16  & 0.35 & 0.3 &1.0\\
  '' & FU &   0.51 $\pm$ 0.10   & 0.31 $\pm$ 0.16    & 0.40 & 0.3 & 1.0 \\
  '' & FU  &    0.50 $\pm$ 0.10   & 0.31 $\pm$ 0.16  & 0.45 & 0.3 &1.0\\
\\
   W($H,K_{\rm S}$) & FU  &    0.56 $\pm$ 0.07\tablenotemark{b}   & 0.50 $\pm$ 0.08\tablenotemark{b}  & \ldots& 0.3 & 1.0 \\
  '' & FU  &    0.56 $\pm$ 0.12   & 0.52 $\pm$ 0.16  & 0.35 & 0.3 &1.0\\
  '' & FU &   0.56 $\pm$ 0.14   & 0.51 $\pm$ 0.15    & 0.40 & 0.3 & 1.0 \\
  '' & FU  &    0.56 $\pm$ 0.13   & 0.52 $\pm$ 0.17  & 0.45 & 0.3 &1.0\\
\enddata    
\tablecomments{The errors on the 
relative distances were estimated by accounting for the uncertainties both in the 
coefficients and in the dispersion of the individual PW relations. }
\tablenotetext{a}{$\Delta\mu_{short} = a1_s + b1_s*x_s - (a2_s+b2*x_s) $ and $\Delta\mu_{long} = a1_l +b1_l*x_l - (a2_l+b2*x_l) $, 
where the index 1 refers to the SMC linear regressions, while the index 2 refers to the LMC linear regressions.}
\tablenotetext{b}{Relative distance moduli estimated using the PW relations on the entire range of periods. $\Delta\mu_{short} $ here is the relative distance moduli obtained at $\log P = x_s$, while $\Delta\mu_{long} $ is obtained at $\log P = x_l$ }
\end{deluxetable} 																  

\begin{deluxetable}{llrlccc}
\tabletypesize{\scriptsize}
\tablewidth{0pt}

\tablecaption{Optical and NIR PW relations available in the literature.}
\label{tab}
	
\tablehead{
\colhead{W($\lambda_2$,$\lambda_1$)}&
\colhead{Mode}&
\colhead{a}&
\colhead{b}&
\colhead{$\sigma$}&
\colhead{galaxy}&
\colhead{Reference}
}
\startdata
W($J,K_{\rm S}$)  & FU (229) & -2.65 $\pm$ 0.02 &  -3.34 $\pm$  0.03 & 0.10  & MW & N12\\
W($J,K_{\rm S}$)  & FU (70) &  -2.52 $\pm$ 0.12 &   -3.44 $\pm$ 0.09 & 0.23  & MW & S11a\\
W($V,I$)  & FU (70) &  -2.70 $\pm$ 0.15 &   -3.26 $\pm$ 0.11 & 0.26  & MW & S11a\\
W($V,I$)  & FU (10) &  -2.48 $\pm$ 0.15 &   -3.37  $\pm$ 0.12 & 0.11  & MW & B07\\
W($V,K_{\rm S}$)  & FU (10) &  -2.60 $\pm$ 0.07 &   -3.325  $\pm$ 0.014 & 0.08  & LMC &R12\\
\enddata
\end{deluxetable} 											  

\clearpage
\begin{figure}[!ht]
\begin{center}
\label{fig1}
\includegraphics[width=0.90\textwidth]{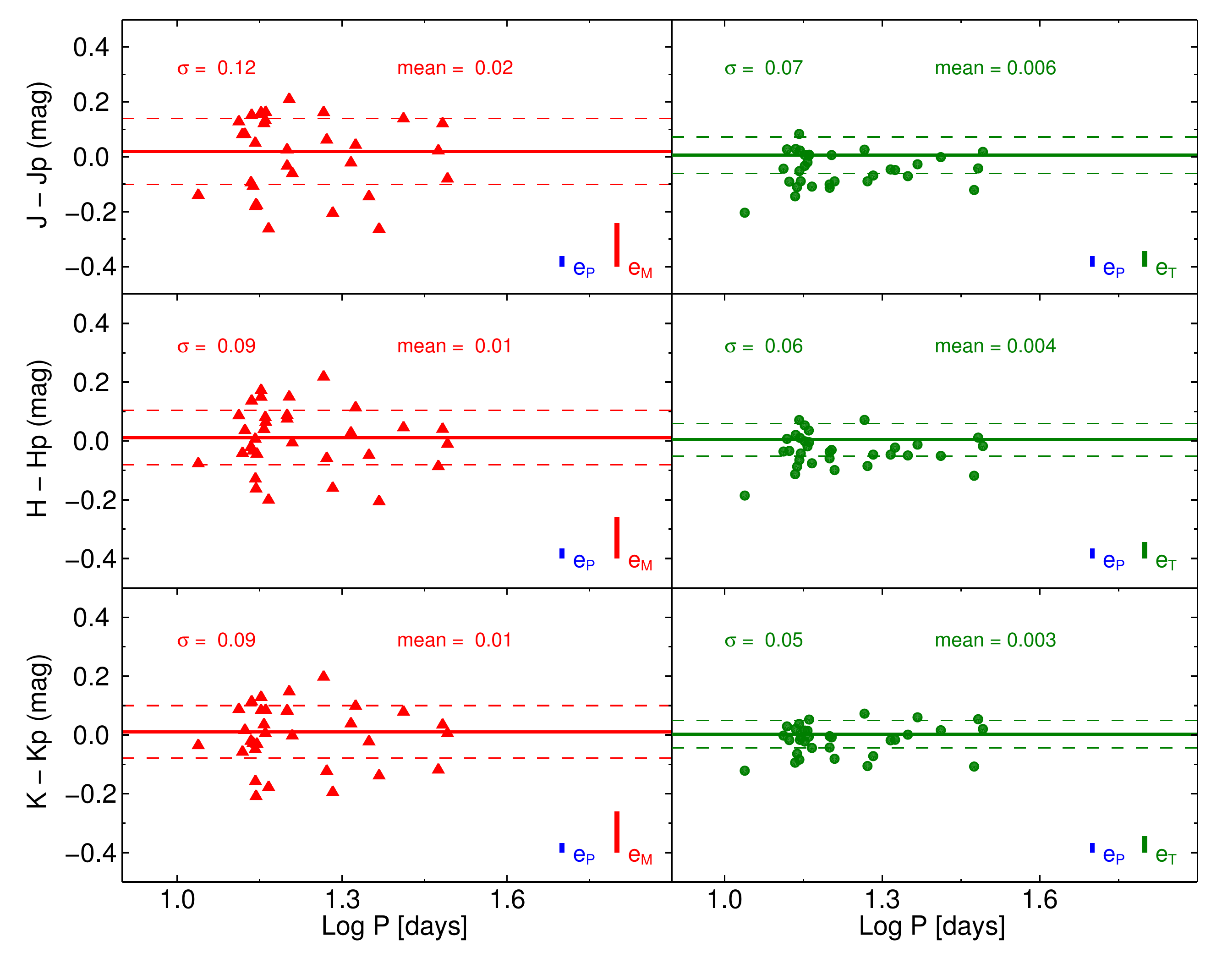}
\caption{Left: difference between the $J$ (top), $H$ (middle) and $K_{\rm S}$ (bottom) 
Cepheid single epoch magnitudes and the true $Jp, Hp, Kp_{\rm S}$ mean magnitudes 
from P04.  For each band is plotted the mean difference (solid red line) and the 
dispersion ($\sigma$, dashed red line). The blue and the red bars in the right 
corner show the typical photometric error associated to the true mean magnitudes
($e_P$) and the error of IRSF magnitudes ($e_M$, photometric error plus the semi-amplitude).
Right: Same as the left, but difference is between the $J$ (top), $H$ (middle) and $K_{\rm S}$
 (bottom) mean magnitudes obtained by applying the template to 
single epoch magnitudes and the true $Jp, Hp, Kp_{\rm S}$ mean magnitudes from P04.
The solid and dashed green lines have the same meaning, while the green bar shows 
the total error ($e_T$, photometric error plus error of the template light curve).    
}
\end{center}
\end{figure}

\clearpage
\begin{figure}[!ht]
\begin{center}
\label{fig2}
\includegraphics[width=0.90\textwidth]{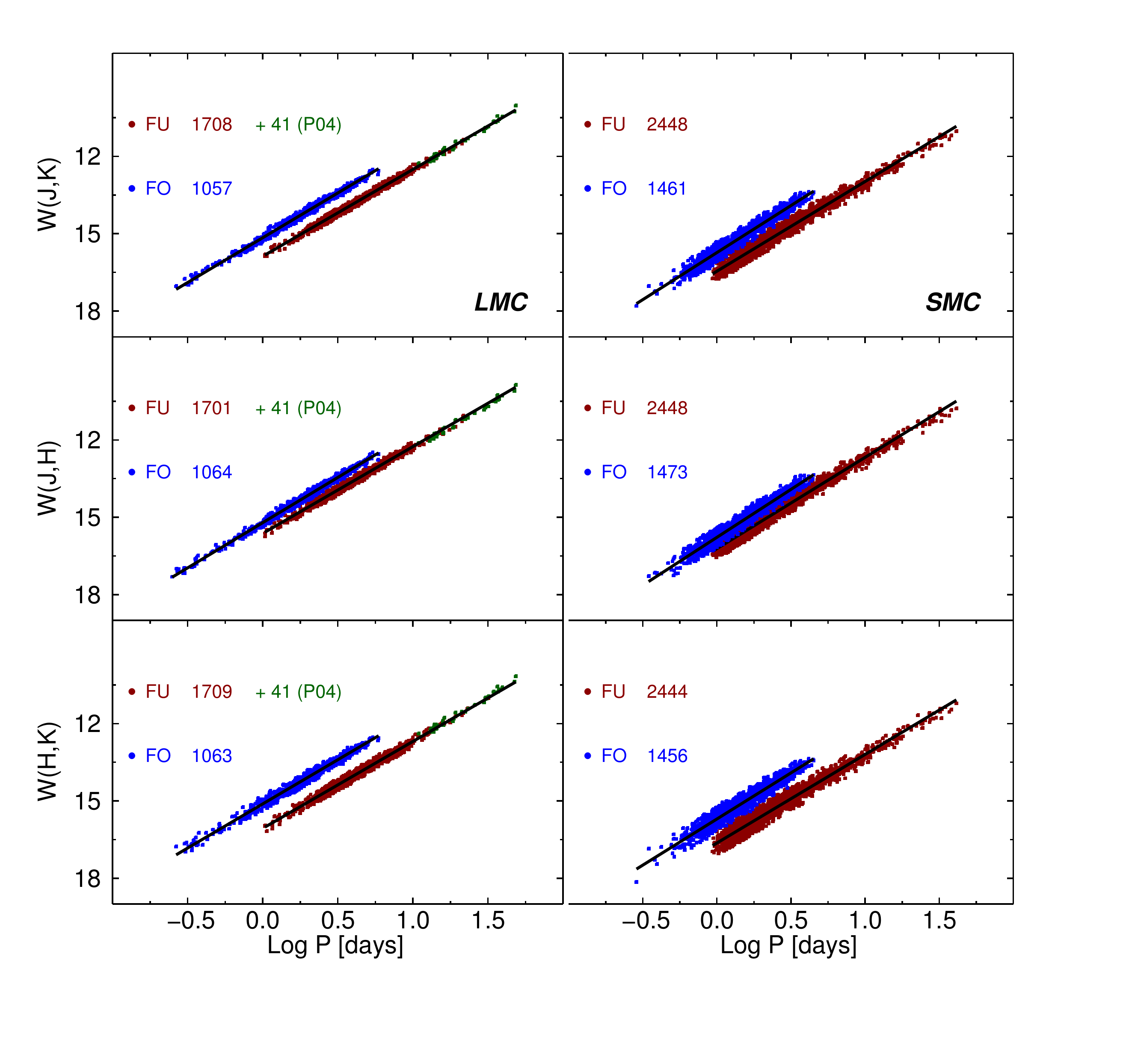}
\vspace*{1truecm}
\caption{NIR Period-Wesenheit relations for LMC (left) and SMC (right) Cepheids.
Red dots display IRSF mean magnitudes for FU pulsators, while green dots are the 
mean magnitude for 41 Cepheids by P04. Blue dots show the FO pulsators. 
The solid lines show the linear fits.
}
\end{center}
\end{figure}

\begin{figure}[!ht]
\begin{center}
\label{fig3}
\includegraphics[height=0.90\textwidth,width=0.75\textwidth]{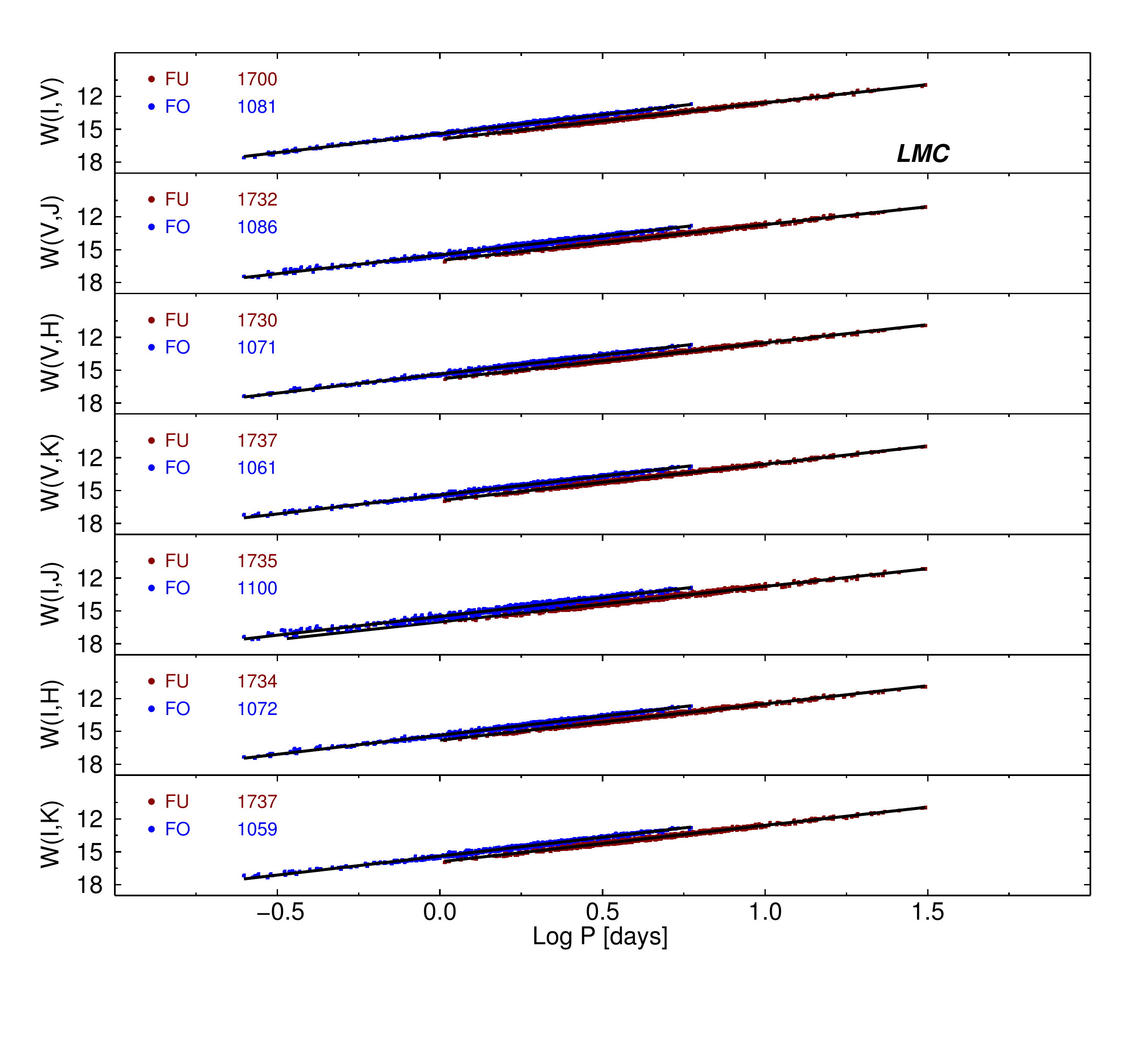}
\caption{Optical-NIR Period-Wesenheit relations for LMC Cepheids. 
Symbols and lines are the same as Figure~2. 
}

\end{center}
\end{figure}

\begin{figure}[!ht]
\begin{center}
\label{fig4}
\includegraphics[height=0.90\textwidth,width=0.75\textwidth]{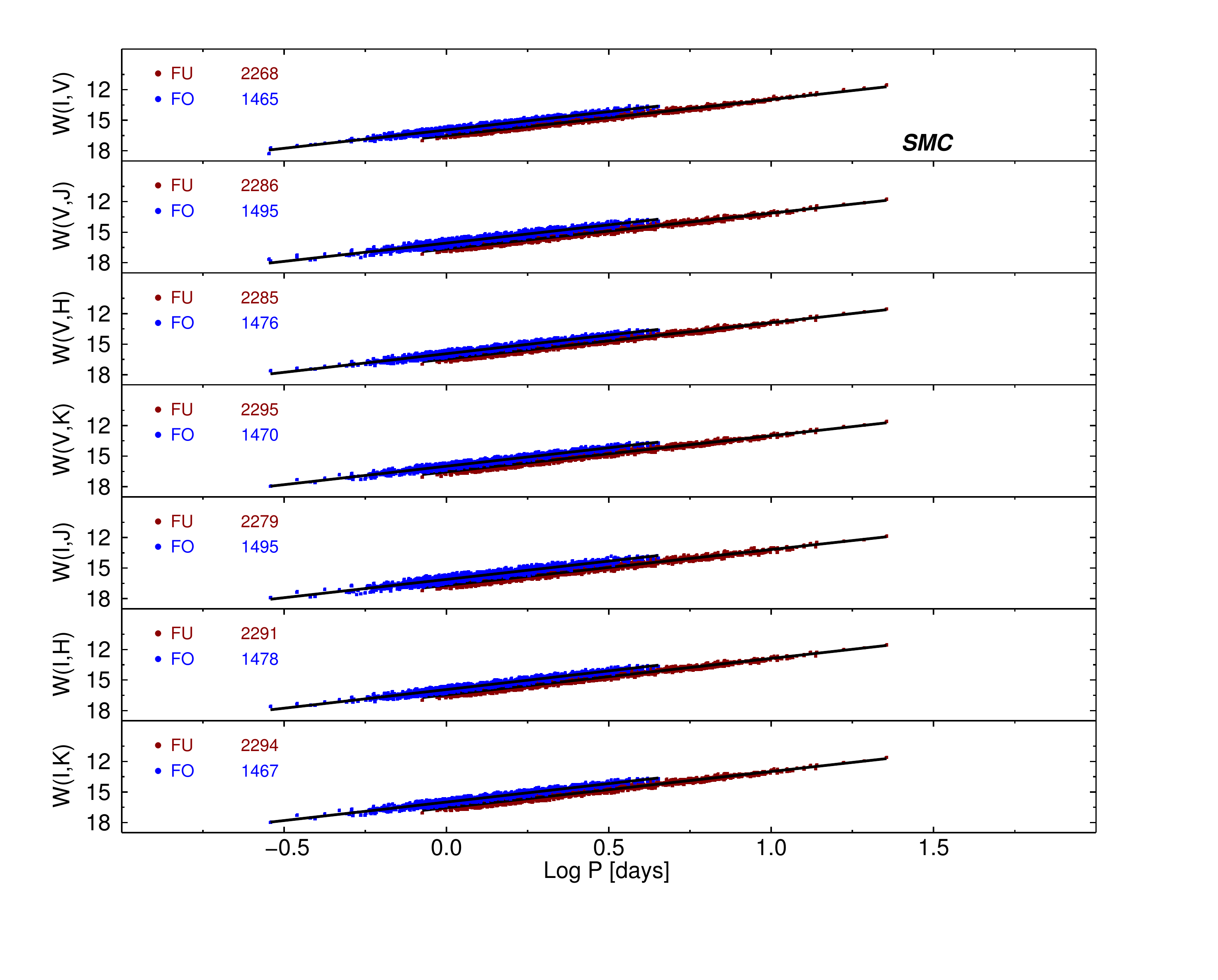}
\caption{Same as Figure~3, but for SMC Cepheids.}
\end{center}
\end{figure}

\begin{figure}[!ht]
\begin{center}
\label{fig5}
\includegraphics[height=0.90\textwidth,width=0.75\textwidth]{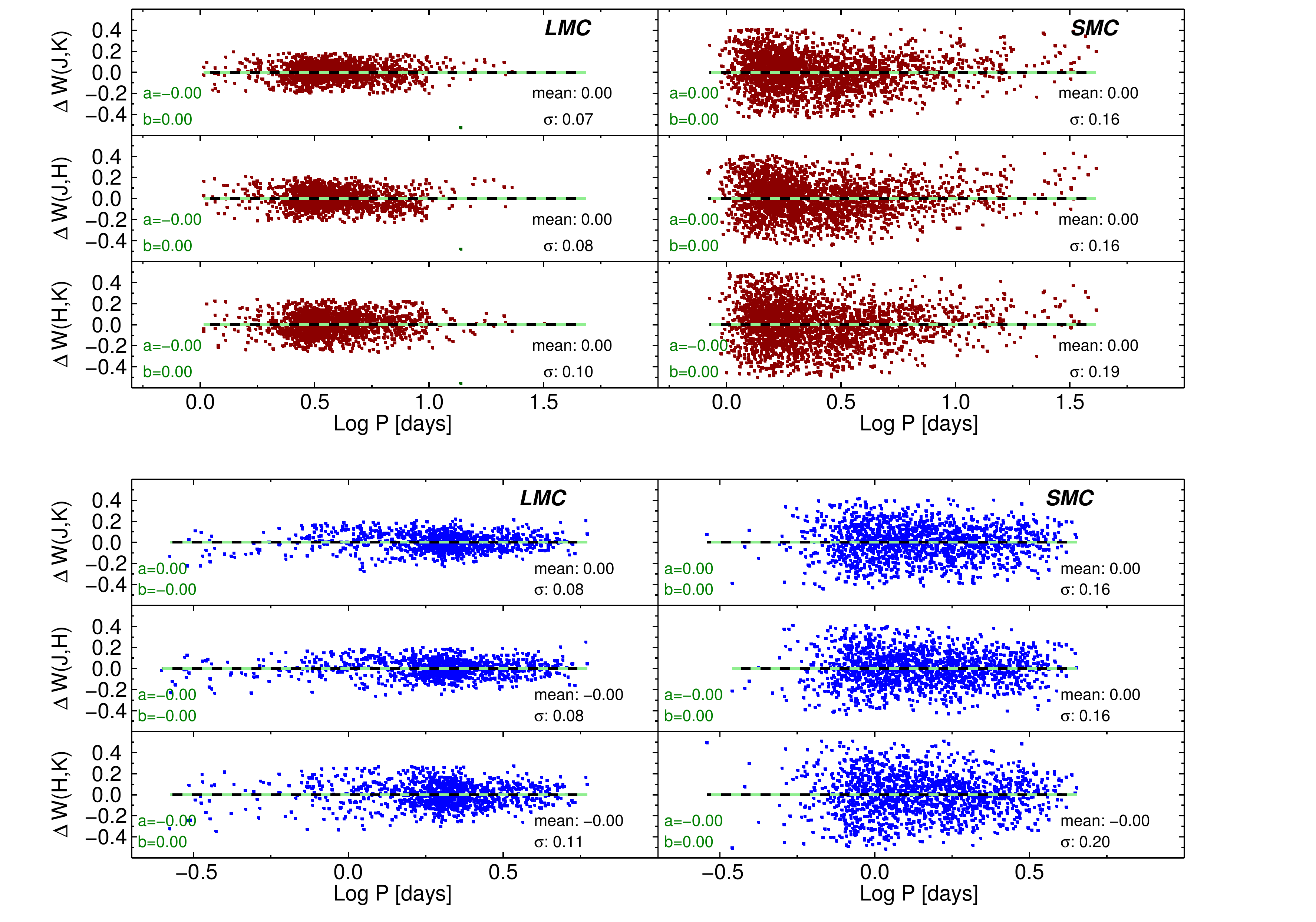}
\caption{Top -- Difference between individual NIR Wesenheit mean magnitudes of 
LMC (left) and SMC (right) Cepheids and the PW relations. 
The linear fit of the residuals is also overplotted (dashed green line). 
The weighted means 
and the intrinsic dispersions are also labeled. 
Bottom -- Same as the top but for FO Cepheids.}
\end{center}
\end{figure}

%
\begin{figure}[!ht]
\begin{center}
\label{fig6}
\includegraphics[height=0.80\textwidth,width=0.80\textwidth]{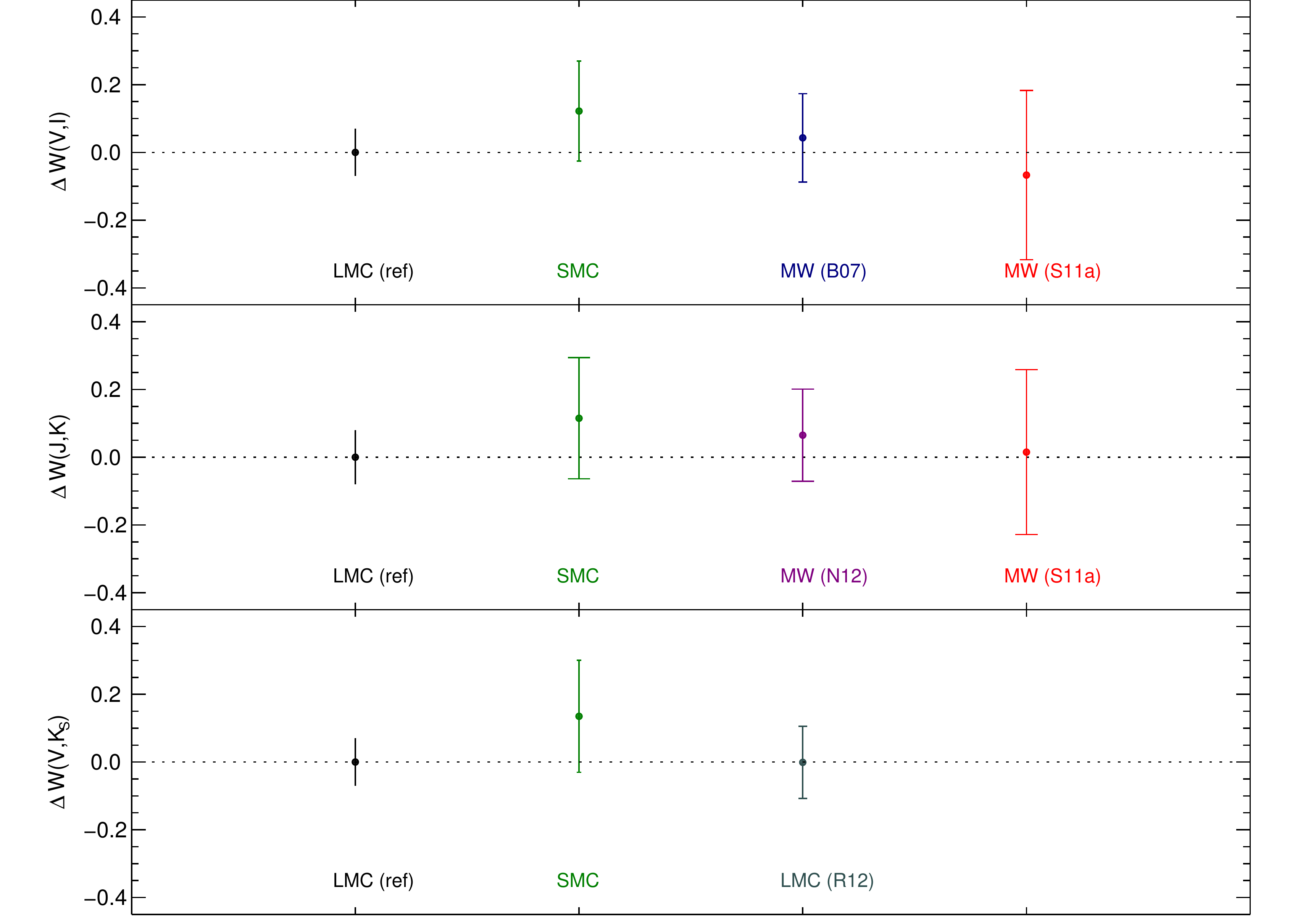}
\caption{Top -- Difference in the slope between the PW($V$,$I$) relation for 
the LMC (black), SMC (green) and similar slopes for Galactic Cepheids 
provided by B07 (blue) and by S11a (red). The vertical error bars display the 
dispersion of the different PW relations.
Middle -- Same as the top, but for the PW($J$,$K_{\rm S}$) relations. The slopes 
for Galactic Cepheids were provided by N12 (purple) and by S11a (red).
Bottom -- Same as the top, but for the PW($V$,$K_{\rm S}$) relations. The slope 
for LMC Cepheids was provided by R12 (grey).  
}
\end{center}
\end{figure}

\end{document}